# Stability of the genetic code and optimal parameters of amino acids

V. R. Chechetkin[a*] and V.V. Lobzin[b]

[a]*Theoretical Department of Division for Perspective Investigations, Troitsk Institute of Innovation and Thermonuclear Investigations (TRINITI), Troitsk, 142190 Moscow Region, Russian Federation*

[b]*School of Physics, University of Sydney, Sydney, NSW 2006, Australia*

**Abstract**

The standard genetic code is known to be much more efficient in minimizing adverse effects of misreading errors and one-point mutations in comparison with a random code having the same structure, i.e. the same number of codons coding for each particular amino acid. We study the inverse problem, how the code structure affects the optimal physico-chemical parameters of amino acids ensuring the highest stability of the genetic code. It is shown that the choice of two or more amino acids with given properties determines unambiguously all the others. In this sense the code structure determines strictly the optimal parameters of amino acids. In the code with the structure of the standard genetic code the resulting values for hydrophobicity obtained in the scheme "leave one out" and in the scheme with fixed maximum and minimum parameters correlate significantly with the natural scale. This indicates the co-evolution of the genetic code and physico-chemical properties of amino acids.

*Keywords:* Genetic code; Error-minimization theory; Mutational and translational stability; Local stability of the genetic code; Co-evolution of the genetic code and physico-chemical properties of amino acids

**1. Introduction**

The genetic code codes for amino acids, which are the building blocks of a variety of proteins. At the present stage of molecular evolution both the genetic code and physico-chemical properties of amino acids should be treated as given and one may only guess how the genetic code evolved and how amino acids were recruited (Woese, 1965; Freeland et al., 2000; Trifonov, 2000; Wu et al., 2005; Copley et al., 2005; Zhu and Freeland, 2006; Stoltzfus and Yampolsky, 2007; Di Giulio, 2008; Novozhilov and Koonin, 2009; for reviews see Knight et al., 2001; Di Giulio, 2005; Koonin and Novozhilov, 2009). The mode of coding must mitigate the adverse effects of mutations and misreading during translation. This is the keystone of the error minimization theory (Alff-Steinberger, 1969; Haig and Hurst, 1991; Freeland and Hurst, 1998). The superior robustness of the standard genetic code over the huge majority of random codes has been evidenced in the numerous publications (Haig and Hurst, 1991; Goldman, 1993; Freeland and Hurst, 1998; Ardell, 1998; Gilis et al., 2001; Luo and Li, 2002; Goodarzi et al., 2004, 2005; Sella and Ardell, 2006; Novozhilov et al., 2007). The error minimization theory permits to prove some rigorous results for the binary codes and the binary subdivisions of amino acid properties reflecting the coarse-grained structure of the standard genetic code (Chechetkin, 2003, 2006). The theory provides also the opportunity of complete enumeration of the small modifications of the standard genetic code and its variants (Novozhilov et al., 2007; Chechetkin and Lobzin, 2009). The approach to the stability problem used in the cited papers was based on the set of amino acids with fixed natural physico-chemical properties and commonly on the fixed number of codons in sets coding for each particular amino acid. The typical questions are: how does a random distribution of codons over coding sets affect the stability of the genetic code and what is the chance to obtain the improvement of the code stability by such occasional distributions?

Our approach is based on the extension of the error minimization theory. In this paper we will consider the inverse problem and try to answer the question: which values of parameters of amino acids ensure the highest stability the genetic code? The consideration proves that the structure of the genetic code strictly determines the optimal parameters of amino acids. In particular, the choice of two or more amino acids with given properties determine unambiguously all the others. In the code with the structure of the standard genetic code

---

*Corresponding author. *E-mail addresses:* chechet@biochip.ru ; vladimir_chechet@mail.ru

the resulting values for hydrophobicity obtained in the scheme "leave one out" and in the scheme with fixed maximum and minimum parameters correlate significantly with the natural scale. This means the co-evolution of the code structure and physico-chemical parameters of recruited amino acids. If, for example, the primordial genetic code has been developed by a half of existing amino acids (Higgs and Pudritz, 2009; Novozhilov and Koonin, 2009), then subsequent recruiting of amino acids would depend on the developed code structure. The same concerns the artificial modifications of the genetic code (Wang and Schultz, 2005; Wang et al., 2009). The additional mechanism of adaptation of the physico-chemical parameters is related to the numerous post-translational modifications of amino acids (Black and Mould, 1991; Lewin, 2000). This mechanism may be relevant to our consideration, but the overall impact of post-translational modifications is difficult to assess.

The plan of the paper is as follows. In Section 2 we present the general scheme and derive the equations for the optimal parameters of amino acids. The scheme allows treating stop-codons as coding for a virtual amino acid and to prescribe the corresponding optimal parameters to the stop-codons as well. Different hydrophobicity scales obtained by solutions of the equations for the optimal parameters of amino acids are compared to the natural scale in Section 3. The effects of the hydrophobicity scales on the local stability of the genetic code and on its binary block structure are also considered in this section. The summary of results and possible consequences from them are discussed in concluding Section 4.

## 2. Stability function and optimal parameters of amino acids

### 2.1. Stability function

The robustness of a code is assessed with stability function

$$\varphi(a(c)) = \sum_c f(c) \sum_{c'} p(c'|c) d(a(c), a(c')), \quad (1)$$

where codon $c$ codes for amino acid $a$ or stop-codon, $f(c)$ is the frequency of codon $c$, $p(c'|c)$ corresponds to the conditional probability for codons $c$ and $c'$, and $d(a(c), a(c'))$ is the cost related to exchange of amino acids $a(c)$ and $a(c')$. The most robust code corresponds to the minimum of the stability function (1).

Below we use a variant with $f(c) = $ const and the cost

$$d(a(c), a(c')) = (P(a(c)) - P(a(c')))^2, \quad (2)$$

where $P(a(c))$ is a physico-chemical characteristic of amino acid. The unbiased conditional probability follows the form proposed by Haig and Hurst (1991), $p(c'|c) = 1/9$ if codons $c$ and $c'$ differ by one-nucleotide replacement and $p(c'|c) = 0$ otherwise, whereas the biased conditional probability corresponds to that proposed by Freeland and Hurst (1998),

$$p(c'|c) = \begin{cases} 1/N & \text{if } c' \text{ and } c \text{ differ in the 3rd} \\ & \text{base only} \\ 1/N & \text{if } c' \text{ and } c \text{ differ in the 1st} \\ & \text{base only and cause a transition} \\ 0.5/N & \text{if } c' \text{ and } c \text{ differ in the 1st base} \\ & \text{only and cause a transversion} \\ 0.5/N & \text{if } c' \text{ and } c \text{ differ in the 2nd base} \\ & \text{only and cause a transition} \\ 0.1/N & \text{if } c' \text{ and } c \text{ differ in the 2nd base} \\ & \text{only and cause a transversion} \\ 0, & \text{otherwise} \end{cases}$$

Factor $N$ is defined by the normalization of conditional probability $\sum_{c'} p(c'|c) = 1$ and is equal to 5.7 in this model. For brevity these conditional probabilities will be called below HH and FH, respectively.

### 2.2. Optimal parameters of amino acids

The optimal parameters of amino acids should also correspond to the minimum of the stability function. They can be derived by parameter differentiation of the stability function (1) with the cost (2), thereby producing the set of equations

$$\sum_{c \in a} \sum_{c'} p(c'|c)(f(c) + f(c'))(P(a(c)) - P(a(c'))) = 0 \quad (3)$$

Both the stability function (1) and the set (3) are code sensitive, i.e. depend on a subdivision of code by the sets coding for different amino acids or stop-codons as well as on distribution of codons over coding sets. Besides the trivial solution of the set (3), in which all the parameters take the same value, there is a large family of non-trivial solutions. The non-trivial solutions of the set (3) possess the following properties: the choice of two or more amino acids with given parameters $P(a(c))$ determines unambiguously all the others. The fixed chosen parameters for a part of amino acids may be considered as additional constraints imposed on the minimization of the stability function (1) with respect to $P(a(c))$. These solutions depend on the conditional probability $p(c'|c)$, frequency $f(c)$, the code structure, and the choice of amino acids with given parameters. The stop-codons may either be skipped or treated as coding for a virtual additional amino acid with optimal parameters. In the case when $f(c)$ = const and the conditional probability is unbiased, the meaning of solutions becomes especially simple. Any parameter $P(a(c))$ for a given coding set would be equal to the average over parameters obtained by various one-point replacements within given codons. Such rule was used previously for the definition of optimal parameters for stop-codons by Goodarzi et al. (2004).

## 3. Comparison of different hydrophobicity scales

### 3.1. General comparison

In this section the results of calculations for hydrophobicity scale are presented. We used two choices for the given parameters of amino acids which determine the others. The first choice corresponds to the statistical procedure "leave one out" (or jack-knife) (Weir, 1990). The values for 19 amino acids were taken from the natural scale, whereas the remaining unique value was calculated from the corresponding equation in the set (3). The procedure was repeated for all 20 amino acids. Then, the resulting scale was remapped onto the interval (0, 1). In the scheme with included stop-codons the optimal value for the stop-codons was added to the calculations. In the second choice only maximum and minimum parameters for phenylalanine and arginine were fixed ($H$(F) = 1; $H$(R) = 0), while all other parameters were calculated from the set (3) with equations for phenylalanine and arginine being discarded. In the scheme "leave one out" and for subsequent comparison we used the hydrophobicity scale by Black and Mould (1991). The calculations with hydrophobicity scale proposed by Tolstrup et al. (1994) provided similar results, but the correspondence was a bit worse. The calculations for molecular volume did not reveal significant correlations with the counterpart natural scale.

The results are summarized in Table 1. As an example, Fig. 1 illustrates the visual correspondence between the natural scale and the optimal hydrophobicity scales obtained with biased conditional probability, skipped stop-codons and the standard genetic code. Fig. 1 shows clearly the narrowing of the dynamic range for the scales with the optimal parameters in comparison with the natural scale (the tangents of broken lines are less than unity). Interestingly, the inclusion of stop-codons with optimal parameters into the whole scheme enhances the correspondence with the natural scale.

### 3.2. Optimal hydrophobicity scales and local stability of the genetic code

The local stability of the standard genetic code was assessed with reassigned hydrophobicity scales following the technique used by Chechetkin and Lobzin (2009). Small modifications of the code were produced either by picking up a codon from a set coding for a particular amino acid and transposing it to a set coding for a different amino acid and containing a close codon differing from original one only by one-nucleotide replacement (the operation termed "delin") or by the swaps of close codons in different sets. In the scheme with the natural hydrophobicity scale and included stop-codons the parameters for stop-codons were calculated with "leave one out" procedure. The relative robustness of a code obtained under delin or swap modifications of the standard code was assessed via the difference

$$\Delta\varphi = \varphi_{\text{modified}} - \varphi_{\text{standard}}. \qquad (4)$$

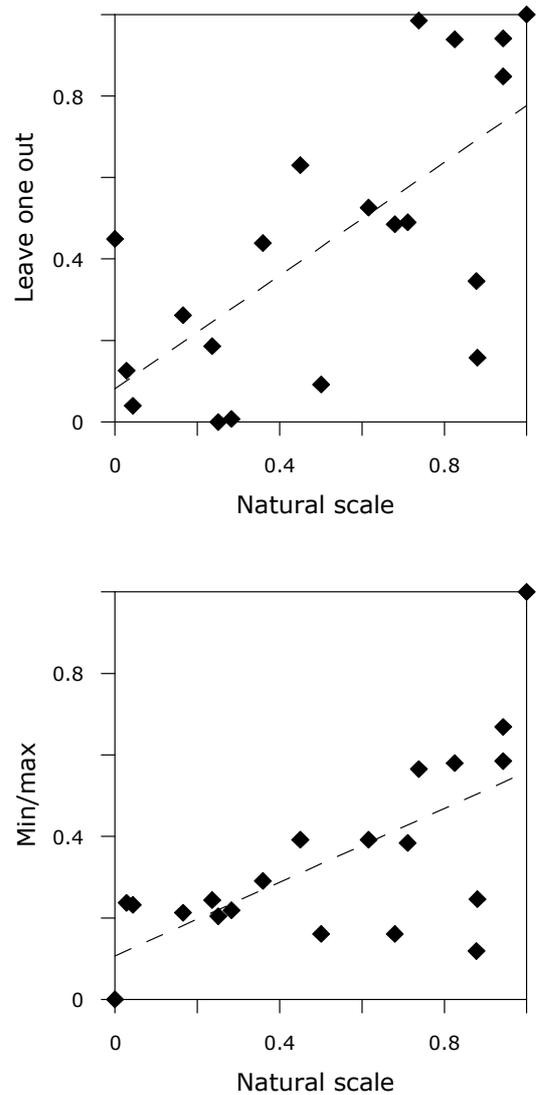

Fig. 1. Comparison between the optimal hydrophobicity scales and the natural scale. The optimal hydrophobicity parameters were calculated with the schemes "leave one out" and min/max (in the latter scheme the minimum and maximum values of hydrophobicity for phenylalanine and arginine are fixed: $H$(R) = 0; $H$(F) = 1). These two hydrophobicity scales were obtained with the biased conditional probability (Freeland and Hurst, 1998; FH) and the standard genetic code with skipped stop-codons. The broken line shows the best linear fit.

The negative variations $\Delta\varphi$ mean that the modifications yield more robust codes in comparison with the standard one. The variations (4) may be conveniently normalized as

$$\Delta\tilde{\varphi} = \Delta\varphi / \sigma(\Delta\varphi), \qquad (5)$$

where $\sigma(\Delta\varphi)$ is the standard deviation calculated for all variations produced by given modifications of the standard code.

The final results are collected in Table 2 and illustrated by Figs. 2 and 3. They prove the enhancement of the local stability of the genetic code with the optimal parameters of amino acids. Although the impact of optimal parameters for the stop-codons on the stability of the genetic code is distinctly less than for that chosen by Chechetkin and Lobzin (2009), their inclusion deteriorates generally the stability for the biased conditional probability, whereas for the unbiased conditional probability the definite trend is absent.

### 3.3. Optimal hydrophobicity scales and local stability of the genetic code

The stability of a code will be improved, if the binary coarse-grained structure exists within a code (Chechetkin, 2003). It is well known that such a structure is inherent to the standard genetic code. In particular, in the binary blocks NRN-NYN the block NRN codes for the amino acids with lower hydrophobicity, whereas the block NYN codes for the amino acids with higher hydrophobicity (Jungck, 1978; Wolfenden et al., 1979; Blalock and Smith, 1984; Taylor and Coates, 1989; Chechetkin, 2003; Wilhelm and Nikolaeva, 2004). The destruction of optimal block structure of the genetic code by delins and swaps diminishes the code stability and should be suppressed. Therefore, the binary block structure imposes certain restrictions on the non-diagonal elements in the matrix of replacements corresponding to delin and swap modifications of the genetic code leading to the negative variations of stability function (Chechetkin and Lobzin, 2009). If hydrophobicity is used as a physico-chemical parameter in the cost given by Eq. (2), among all binary subdivisions NRN-NYN, NWN-NSN, and NMN-NKN (here R = (A, G) and Y = (C, U), W = (A, U) and S = (C, G), K = (G, U) and M = (A, C), and N is any nucleotide), the sum of non-diagonal elements should be the smallest for the subdivision NRN-NYN. The ratio of the sum of non-diagonal elements to the total sum over all matrix elements for delin operations should be significantly lower than the expected value for random codes (about $2/9 \approx 0.22$). The corresponding ratio for swaps must include only one of non-diagonal elements due to strict symmetry for swap matrix. These features are clearly seen in Table 3. The violation of these rules for the scale "leave one out" (see Table 3B) may lead to the partial deterioration of the local stability of the genetic code (cf. Table 2A, columns FH, swaps).

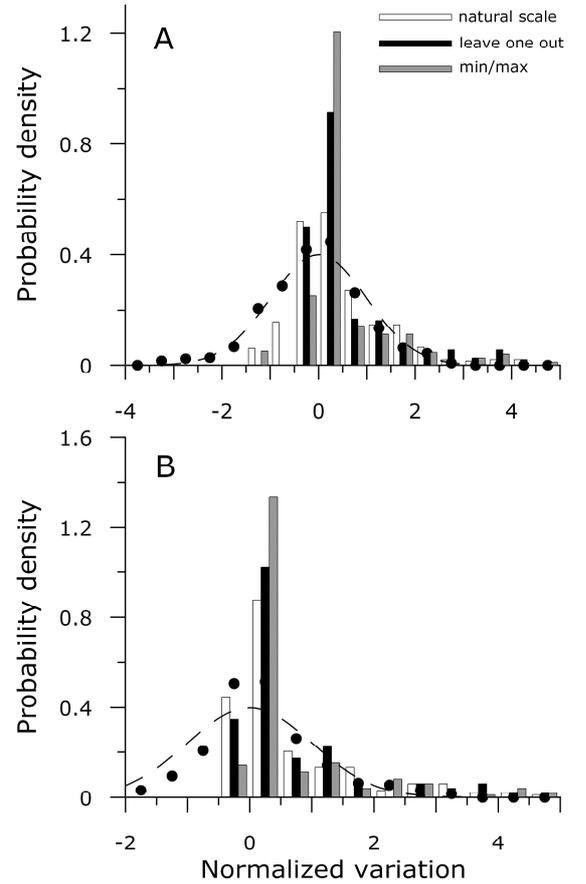

Fig. 2. The dependence on the hydrophobicity scale of normalized variations of stability function produced by delins (A) and swaps (B) in the standard code with skipped stop-codons. The broken line shows the Gaussian distribution. The circles correspond to the distribution of normalized variations for a particular random code with the same number of codons in the sets coding for different amino acids as in the standard code. The distributions were obtained with the biased conditional probability (FH).

On the contrary, the better fulfillment of these rules for the min/max scale leads to the higher local stability.

The separate blocks within the genetic code may be considered as subcodes within the code. The set of equations for amino acid parameters like (3) can be reduced to particular blocks NRN-NYN, NWN-NSN, or NMN-NKN by skipping irrelevant amino acids. The other natural subdivision is related to two classes of aminoacyl-tRNA synthetases.

### 4. Discussion

Our consideration shows that the optimal parameters of amino acids improve the stability of the genetic code. The results obtained for the optimal hydrophobicity and the standard genetic code correlate significantly with the natural scale. This indicates that the stability of the genetic code may be considered as one of the major driving forces during evolution of the genetic code. The codons coding for amino acids providing the highest divergence between natural and optimal values are often reassigned in the variants of the genetic code (see Table 1 and cf., Knight et al., 2001; Chechetkin and Lobzin, 2009). The enhancement of stability for the optimal parameters of amino acids is partially attained at the expense of the narrowing of the dynamic range in the optimal scales (cf. Fig. 1), whereas in the natural scale the values of hydrophobicity are distributed approximately uniformly over the interval (0, 1).

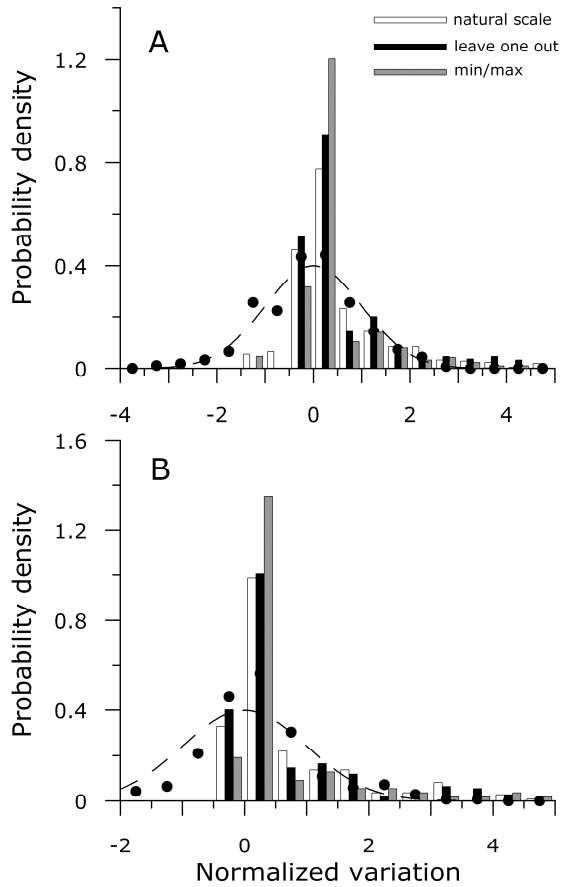

Fig. 3. The dependence on the hydrophobicity scale of normalized variations of stability function produced by delins (A) and swaps (B) in the standard code with included stop-codons. Stop-codons are assumed to code for a virtual amino acid with optimal hydrophobicity. The broken line shows the Gaussian distribution. The circles correspond to the distribution of normalized variations for a particular random code with the same number of codons in the sets coding for different amino acids or stop-codons as in the standard code. The distributions were obtained with the biased conditional probability (FH).

The resulting scale depends on the choice of amino acids with fixed parameters which determine all the others. The complete number of choices of two or more amino acids from the set of twenty amino acids is equal to

$$\sum_{k=2}^{19} C_k^{20} \approx 2^{20} \approx 10^6, \quad (6)$$

where $C_k^n = \dfrac{n!}{k!(n-k)!}$. The corresponding value for the scheme with stop-codons is twice more. This combinatorics is incomparable with the complete number of variants for the genetic code $\sim 3 \times 10^{83}$ and may be enumerated with modern computers. Among the potentially interesting choices of amino acids with given parameters it is worth mentioning the set of ten amino acids which are thought to be the most ancient and formed the primordial genetic code, {Gly, Ala, Asp, Glu, Val, Ser, Ile, Leu, Pro, Thr} (Higgs and Pudritz, 2009; Novozhilov and Koonin, 2009), or amino acids for two classes of aminoacyl-tRNA synthetases. The stricter correspondence of the calculated values for the remaining amino acids in comparison with the counterparts obtained by the arbitrary choices of ten given amino acids from the complete set of twenty amino acids may provide additional arguments in favor of the suggested evolutionary scenario. The same concerns the local stability of the genetic code. The relevant combinatorics is rather large and comprises about $C_{10}^{20} \sim 10^5$ variants, which is beyond our aims in this paper.

The assessment of optimal parameters for conditional probability within the frameworks of the inverse approach is also of basic interest. Unlike physico-chemical parameters, the optimal parameters for conditional probability cannot be found by the proper minimization of stability function (1). Such an optimization directly within scheme (1) would produce a trivial answer with overweighted synonymous replacements. The potential approach to this problem may be based on the dependence of optimal physico-chemical parameters on the conditional probability (see Eq. (3)) and on the search of the closest correspondence between the optimal and natural scales ensuring the highest stability of the genetic code.

The further extension of the error minimization theory and the inverse approach to the stability of the genetic code may provide additional insight to the evolutionary origin of the code. It also yields useful dual addition to the direct stability problem. Basically, it reflects self-consistent co-evolution of the code structure and the properties of amino acids.

### References

Alff-Steinberger, C., 1969. The genetic code and error transmission. Proc. Natl. Acad. Sci. USA 64, 584–591.

Ardell, D.H. 1998. On error minimization in a sequential origin of the standard genetic code. J. Mol. Evol. 47, 1–13.

Black, S.D., Mould, D.R., 1991. Development of hydrophobicity parameters to analyze proteins which bear post or cotranslational modifications. Analyt. Biochem. 193, 72–82.

Blalock, J.E., Smith, E.M. 1984. Hydropathic anti-complementarity of amino acids based on the genetic code. Biochem. Biophys. Res. Commun. 121, 203–207.

Chechetkin, V.R., 2003. Block structure and stability of the genetic code. J. Theor. Biol. 222, 177–188.

Chechetkin, V.R., 2006. Genetic code from tRNA point of view. J. Theor. Biol. 242, 922–934.

Chechetkin, V.R., Lobzin, V.V., 2009. Local stability and evolution of the genetic code. J. Theor. Biol. 261, 643–653.

Copley, S.D., Smith, E., Morowitz, H.J., 2005. A mechanism for the association of amino acids with their codons and the origin of the genetic code. Proc. Natl. Acad. Sci. USA, 102, 4442–4447.

Di Giulio, M., 2005. The origin of the genetic code: theories and their relationships, a review. BioSystems 80, 175–184.

Di Giulio, M., 2008. An extension of the coevolution theory of the origin of the genetic code. Biology Direct 3, 37

Freeland, S.J., Hurst, L.D., 1998. The genetic code is one in a million. J. Mol. Evol. 47, 238–248.

Freeland, S.J., Knight, R.D., Landweber, L.F., Hurst, L.D., 2000. Early fixation of an optimal genetic code. Mol. Biol. Evol. 17, 511–518.

Gilis, D., Massar, S., Cerf, N.J., Rooman, M., 2001. Optimality of the genetic code with respect to protein stability and amino acid frequencies. Genome Biol. 2 (11), 49.1–49.12.

Goldman, N., 1993. Further results on error minimization in the genetic code. J. Mol. Evol. 37, 662–664.

Goodarzi, H., Najafabadi H.S., Hassani K., Nejad, H.A., Torabi, N., 2005. On the optimality of the genetic code, with the consideration of coevolution theory by comparison of prominent cost measure matrices. J. Theor. Biol. 235, 318–325.

Goodarzi, H., Nejad, H.A., Torabi, N., 2004. On the optimality of the genetic code, with the consideration of termination codons. BioSystems. 77, 163–173.

Haig, D., Hurst, L.D., 1991. A quantitative measure of error minimization on the genetic code. J. Mol. Evol. 33, 412–417.

Higgs, P.G., Pudritz, R.E., 2009. A thermodynamic basis for prebiotic amino acid synthesis and the nature of the first genetic code. Astrobiology. 9, 483–490.

Jungck, J.R., 1978. The genetic code as a periodic table. J. Mol. Evol. 11, 211–224.

Knight, R.D., Freeland, S.J., Landweber, L.F., 2001. Rewiring the keyboard: evolvability of the genetic code. Nature Reviews Genetics 2, 49–58.

Koonin, E.V., Novozhilov, A.S., 2009. Origin and evolution of the genetic code: the universal enigma. IUBMB Life 61, 99–111.

Lewin, B., 2000. Genes VII. Oxford University Press, New York.

Luo, L., Li, X., 2002. Coding rules for amino acids in the genetic code: the genetic code is a minimal code of mutational deterioration. Orig. Life Evol. Biosph. 32, 23–33.

Novozhilov, A.S., Koonin, E.V., 2009. Exceptional error minimization in putative primordial genetic codes. Biology Direct 4, 44.

Novozhilov, A.S., Wolf, Y.I., Koonin, E.V., 2007. Evolution of the genetic code: partial optimization of a random code for robustness to translation error in a rugged fitness landscape. Biology Direct 2, 24.

Sella, G., Ardell, D.H., 2006. The coevolution of genes and genetic codes: Crick's frozen accident revisited. J. Mol. Evol. 63, 297–313.

Stoltzfus, A.,Yampolsky, L.Y., 2007. Amino acid exchangeability and the adaptive code hypothesis. J. Mol. Evol. 65, 456–462.

Taylor, F.J., Coates D., 1989. The code within the codons. BioSystems 22, 177–187.

Tolstrup, N., Toftgård, J., Engelbrecht, J., Brunak, S., 1994. Neural network model of the genetic code is strongly correlated to the GES scale of amino acid transfer free energies. J. Mol. Biol. 243, 816–820.

Trifonov, E.N., 2000. Consensus temporal order of amino acids and evolution of the triplet code. Gene 261, 139–151.

Wang, L., Schultz, P.G., 2005. Expanding the genetic code. Angew. Chem. 44, 34–66.

Wang, Q., Parrish, A.R., Wang, L. 2009. Expanding the genetic code for biological studies. Chem. Biol. 16, 323–336.

Weir, B.S., 1990. Genetic Data Analysis: Methods for Discrete Population Genetic Data. Sinauer, Sunderland, Massachusetts.

Wilhelm, T., Nikolaeva, S. 2004. A new classification scheme of the genetic code. J. Mol. Evol. 59, 598–605.

Woese, C.R., 1965. On the evolution of the genetic code. Proc. Natl. Acad. Sci. USA 54, 1546–1552.

Wolfenden, R.V., Cullis, P.M., Southgate, C.C.F., 1979. Water, protein folding, and the genetic code. Science 206, 575–577.

Wu, H.-L., Bagby, S., van den Elsen, J.M.H., 2005. Evolution of the genetic triplet code via two types of doublet codons. J. Mol. Evol. 61, 54–64.

Zhu, W., Freeland, S., 2006. The standard genetic code enhances adaptive evolution of proteins. J. Theor. Biol. 239, 63–70.

**Table 1**

Comparison of different hydrophobicity scales and their Pearson correlations with the natural scale

| Amino acid | | Natural scale | Leave one out | | | | Min/max | | | |
| --- | --- | --- | --- | --- | --- | --- | --- | --- | --- | --- |
| | | | Conditional probability and stop-codons | | | | | | | |
| Three-letter code | One-letter code | | HH, − | FH, − | HH, + | FH, + | HH, − | FH, − | HH, + | FH, + |
| Ala | A | 0.616 | 0.358 | 0.526 | 0.320 | 0.493 | 0.292 | 0.392 | 0.284 | 0.381 |
| Cys | C | 0.680 | 0.527 | 0.485 | 0.489 | 0.462 | 0.357 | 0.161 | 0.339 | 0.161 |
| Asp | D | 0.028 | 0.231 | 0.126 | 0.185 | 0.066 | 0.314 | 0.237 | 0.301 | 0.215 |
| Glu | E | 0.043 | 0.131 | 0.040 | 0.122 | 0.030 | 0.293 | 0.232 | 0.285 | 0.208 |
| Phe | F | 1.000 | 1.000 | 1.000 | 1.000 | 1.000 | 1.000 | 1.000 | 1.000 | 1.000 |
| Gly | G | 0.501 | 0.175 | 0.092 | 0.138 | 0.052 | 0.238 | 0.161 | 0.233 | 0.156 |
| His | H | 0.165 | 0.229 | 0.262 | 0.183 | 0.211 | 0.277 | 0.213 | 0.264 | 0.190 |
| Ile | I | 0.943 | 0.652 | 0.848 | 0.631 | 0.837 | 0.382 | 0.585 | 0.373 | 0.573 |
| Lys | K | 0.283 | 0.000 | 0.007 | 0.000 | 0.000 | 0.256 | 0.219 | 0.252 | 0.197 |
| Leu | L | 0.943 | 0.687 | 0.941 | 0.648 | 0.930 | 0.406 | 0.669 | 0.390 | 0.653 |
| Met | M | 0.738 | 0.775 | 0.985 | 0.762 | 0.984 | 0.319 | 0.565 | 0.310 | 0.553 |
| Asn | N | 0.236 | 0.250 | 0.186 | 0.206 | 0.130 | 0.311 | 0.244 | 0.297 | 0.222 |
| Pro | P | 0.711 | 0.260 | 0.490 | 0.216 | 0.455 | 0.256 | 0.384 | 0.247 | 0.373 |
| Gln | Q | 0.251 | 0.069 | 0.000 | 0.065 | 0.053 | 0.252 | 0.204 | 0.248 | 0.183 |
| Arg | R | 0.000 | 0.502 | 0.449 | 0.470 | 0.419 | 0.000 | 0.000 | 0.000 | 0.000 |
| Ser | S | 0.359 | 0.506 | 0.439 | 0.473 | 0.402 | 0.298 | 0.291 | 0.288 | 0.282 |
| Thr | T | 0.450 | 0.402 | 0.630 | 0.367 | 0.604 | 0.274 | 0.392 | 0.266 | 0.381 |
| Val | V | 0.825 | 0.704 | 0.939 | 0.686 | 0.934 | 0.385 | 0.580 | 0.375 | 0.568 |
| Trp | W | 0.878 | 0.304 | 0.346 | 0.297 | 0.355 | 0.236 | 0.119 | 0.240 | 0.133 |
| Tyr | Y | 0.880 | 0.225 | 0.158 | 0.240 | 0.272 | 0.426 | 0.246 | 0.382 | 0.206 |
| Ter | * | – | – | – | 0.422 | 0.504 | – | – | 0.284 | 0.176 |
| Correlation coefficients | | | 0.603 | 0.678 | 0.622 | 0.724 | 0.558 | 0.652 | 0.551 | 0.663 |

HH and FH denote unbiased and biased conditional probabilities, respectively. Plus and minus symbols correspond to the codes with included and skipped stop-codons. The characteristic significant values for Pearson correlation coefficients are 0.444 ($Pr = 0.05$) and 0.561 ($Pr = 0.01$).

**Table 2**

Characteristics of negative normalized variations of stability function for modifications of the standard code with different hydrophobicity scales for amino acids

A. Fraction of normalized variations with $\Delta\varphi < 0$

| Operation | Total | Natural scale | | Leave one out | | Min/max | | Stop-codons |
|---|---|---|---|---|---|---|---|---|
| | | HH | FH | HH | FH | HH | FH | |
| Delin | 438 | 0.37 | 0.29 | 0.37 | 0.26 | 0.23 | 0.18 | Yes |
| Delin | 392 | 0.37 | 0.28 | 0.34 | 0.25 | 0.24 | 0.15 | No |
| Swap | 219 | 0.22 | 0.16 | 0.20 | 0.20 | 0.10 | 0.10 | Yes |
| Swap | 196 | 0.22 | 0.16 | 0.23 | 0.17 | 0.11 | 0.07 | No |

For random codes, the expected fraction for $\Delta\varphi < 0$ is 0.5. Parameters for stop-codons in the natural scale are taken from the scheme "leave one out".

B. Relative fraction of normalized variations with $\Delta\varphi < -1$

| Operation | Total | Natural scale | | Leave one out | | Min/max | | Stop-codons |
|---|---|---|---|---|---|---|---|---|
| | | HH | FH | HH | FH | HH | FH | |
| Delin | 438 | 0.18 | 0.09 | 0.12 | 0.00 | 0.16 | 0.13 | Yes |
| Delin | 392 | 0.08 | 0.00 | 0.14 | 0.00 | 0.17 | 0.17 | No |
| Swap | 219 | 0.00 | 0.00 | 0.00 | 0.00 | 0.00 | 0.00 | Yes |
| Swap | 196 | 0.00 | 0.00 | 0.00 | 0.00 | 0.10 | 0.00 | No |

Relative fraction with $\Delta\varphi < -1$ is defined with respect to the number of variations with $\Delta\varphi < 0$. For random codes, the expected fraction for $\Delta\varphi < -1$ is 0.32.

**Table 3**

Representation of replacements in the second codon position leading to the negative variations of stability function in terms of binary codes. Data for different hydrophobicity scales correspond to the biased conditional probability (FH) and to the code without stop-codons.

A. Natural scale

Delins

|   | R  | Y  |   | W  | S  |   | K  | M  |
|---|----|----|---|----|----|---|----|----|
| R | 64 | 5  | W | 65 | 5  | K | 43 | 12 |
| Y | 5  | 37 | S | 7  | 34 | M | 4  | 52 |

Swaps

|   | R  | Y  |   | W  | S  |   | K  | M  |
|---|----|----|---|----|----|---|----|----|
| R | 25 | 3  | W | 17 | 9  | K | 7  | 8  |
| Y | 3  | 3  | S | 9  | 5  | M | 8  | 16 |

B. Leave one out

Delins

|   | R  | Y  |   | W  | S  |   | K  | M  |
|---|----|----|---|----|----|---|----|----|
| R | 53 | 3  | W | 58 | 2  | K | 47 | 5  |
| Y | 2  | 40 | S | 2  | 36 | M | 4  | 42 |

Swaps

|   | R  | Y  |   | W  | S  |   | K  | M  |
|---|----|----|---|----|----|---|----|----|
| R | 22 | 2  | W | 20 | 2  | K | 20 | 4  |
| Y | 2  | 10 | S | 2  | 12 | M | 4  | 10 |

C. Min/max

Delins

|   | R  | Y  |   | W  | S  |   | K  | M  |
|---|----|----|---|----|----|---|----|----|
| R | 27 | 0  | W | 19 | 0  | K | 42 | 4  |
| Y | 0  | 32 | S | 4  | 36 | M | 0  | 13 |

Swaps

|   | R | Y |   | W | S |   | K  | M |
|---|---|---|---|---|---|---|----|---|
| R | 9 | 0 | W | 5 | 2 | K | 12 | 2 |
| Y | 0 | 5 | S | 2 | 7 | M | 2  | 0 |